\documentclass[a4 paper,12pt]{article}
\usepackage{amsmath}
\usepackage{amsfonts}
\usepackage{amssymb}
\usepackage{color}
\usepackage{graphicx}
\usepackage{array}
\usepackage{multirow}
\usepackage{setspace}
\usepackage[margin=1 in]{geometry}
\usepackage{float}
\usepackage[utf8]{inputenc}
\usepackage{xcolor}
\newcolumntype{P}[1]{>{\centering\arraybackslash}p{#1}}
\newcolumntype{M}[1]{>{\centering\arraybackslash}m{#1}}
\usepackage{cite}
\usepackage{hyperref}
\usepackage{authblk}
\begin{document}
%\maketitle
\begin{center}
{\bf\large{FLPR Model: Supervariable approach and a model for Hodge theory} }

\vskip 1.5 cm

{\sf{ \bf Ansha S. Nair$^{a}$, R. Kumar$^{b}$ and Saurabh Gupta$^{a}$}}\\
\vskip .1cm
{\it $^{a}$Department of Physics, National Institute of Technology Calicut,\\ Kozhikode - 673 601, Kerala, India}\\
{\it $^{b}$Department of Physics, Siksha Bhavana,\\
Visva-Bharati, Santiniketan, Bolpur - 731 235, West Bengal, India}
\vskip .15cm
{E-mails: {\tt anshsuk8@gmail.com, raviphynuc@gmail.com, saurabh@nitc.ac.in}}
\end{center}
\vskip 1cm
\noindent
\textbf{Abstract:} 
We procure the complete set of (anti-)BRST as well as (anti-)co-BRST symmetry transformations of the Friedberg-Lee-Pang-Ren (FLPR) model within the framework of the supervariable approach. Additionally, we capture the nilpotency and absolute anti-commutativity aspects of these symmetries, along with the invariance of the Lagrangian in terms of  translational generators along the Grassmannian directions in the context of the supervariable approach. The anti-commutator of the (anti-)BRST and (anti-)co-BRST symmetries generates a novel bosonic symmetry that retains the ghost part of the Lagrangian invariant. Furthermore, we demonstrate that the system under consideration respects the ghost
scale and discrete symmetries, in addition to other symmetries. We show that the generators of all these symmetries cling to the algebra obeyed by the de Rham cohomological operators of differential geometry - thus, the FLPR model presents a toy model for the Hodge theory.


\clearpage

\section{Introduction}
Gauge symmetries play a pivotal role in making the fundamental theories of nature comprehensible. These gauge theories are characterized by the presence of first-class constraints which are the generators of the gauge symmetries~\cite{a,1}. The Becchi-Rouet-Stora-Tyutin (BRST) formalism is one of the sophisticated ways of covariant quantization of gauge theories. Based on the BRST formalism, we rewrite a  gauge invariant theory as a quantum theory that possesses off-shell nilpotent and absolutely anti-commuting global symmetries known as (anti-)BRST symmetries~\cite{2,3,4,5,b}. This quantum theory respects another set of off-shell nilpotent and absolutely anti-commuting symmetries called (anti-)co-BRST symmetries. The (anti-)BRST symmetry transformations leave the kinetic terms invariant in the Lagrangian whereas the gauge-fixing term remains invariant under the (anti-)co-BRST transformations.

The geometrical origin and interpretation  of (anti-)BRST and (anti-)co-BRST transformations have been clearly elucidated using the supervariable approach~\cite{6,7}. In this approach, a $D$-dimensional gauge theory is generalized onto a $(D,2)$-dimensional theory parametrized by an additional pair of Grassmannian superspace variables $(\eta$, $\bar{\eta})$. By exploiting the (dual-)horizontality conditions together with the gauge invariant [or (anti-)BRST] restrictions and (anti-)co-BRST restrictions, we derive the  proper (anti-)BRST and (anti-)co-BRST symmetry transformations of the system. A wide variety of gauge systems such as the Christ-Lee model, rigid rotor, vector Schwinger model, particle on a torus knot and FLRW model have been studied within the framework of the supervariable approach~\cite{8,9,10,11,c}.

The Friedberg-Lee-Pang-Ren (FLPR) model is a $U(1)$ gauge invariant solvable model of a single non-relativistic particle of unit mass with the characteristics of Gribov ambiguity~\cite{12}. This model is endowed with two first-class constraints \textit{\`{a} la} Dirac formalism. A delineated BRST analysis of the model, by summing over all Gribov copies, was carried out in~\cite{13}.  A  quantization technique using physical projector operators, which abstain from the need of gauge fixing, was explored for the FLPR model in~\cite{14}. Recently, we performed Faddeev-Jackiw quantization and BRST analysis of this model using an admissible gauge condition~\cite{15}. Moreover, the presence of (anti-)co-BRST symmetries, in Cartesian coordinates, was demonstrated in~\cite{16}.

Our primary motive for the present endeavor is to derive the (anti-)BRST as well as (anti-)co-BRST symmetry transformations of the FLPR model  by employing the supervariable approach. We also intend to provide a geometrical meaning to these fermionic symmetries. Second, we aspire to reveal the existence of novel symmetries of this model. Finally, yet importantly, we wish to establish that the algebra of these symmetry transformations (and their conserved charges), in the operator form, is analogous to the algebra obeyed by the de Rham cohomological operators of differential geometry. Thus, our aim is to establish the FLPR model as a toy model for the Hodge theory.

The structure of our manuscript is organised as follows. Section~\ref{S2} briefly discusses the gauge invariant theory of the FLPR model. We also provide an account of the off-shell nilpotent and absolutely anti-commuting (anti-)BRST and (anti-)co-BRST symmetries  and their corresponding conserved charges. In section~\ref{S3}, we derive the (anti-)BRST symmetries within the framework of supervariable approach and provide their geometrical interpretation in terms of the Grassmannian translational generators. A similar analysis is being performed in the case of (anti-)co-BRST symmetries in section~\ref{S4}. In our subsequent section~\ref{S5}, we recapitulate the off-shell nilpotency and absolute anti-commutativity of the (anti-)BRST as well as (anti-)co-BRST transformations in terms of the translational generators along the Grassmannian directions. Moreover, we capture the (anti-)BRST as well as (anti-)co-BRST invariance of the Lagrangian within the framework of supervariable approach. Our section~\ref{S6} is devoted to the derivation of bosonic symmetry transformations. The continuous ghost scale and discrete symmetries are listed in section~\ref{S7}. In section~\ref{S8}, we demonstrate that the generators corresponding to the continuous symmetry transformations adhere to an algebra that is obeyed by the de Rham cohomological operators of the differential geometry. Finally, we summarize our results  in section~\ref{S9}.

\section{Prelimineries: Novel Symmetries}\label{S2}
The first-order gauge invariant Lagrangian describing the dynamics of FLPR model, in polar coordinates, is given by \cite{15}
\begin{equation}\label{Lf}
L_{f}=P_{r}\,\dot{r} + P_{\theta}\,\dot{\theta} + P_{z} \,\dot{z}-\frac{1}{2}\,P_{r}^{2} - \frac{1}{2r^{2}}\,P_{\theta}^{2} - \frac{1}{2}\,P_{z}^{2}- \zeta \big(P_z + g\,P_\theta \big) - V(r),
\end{equation}
where $P_{r}$, $P_{\theta}$, $P_{z}$ are canonical momenta corresponding to the generalised coordinates ${r}$, ${\theta}$ and ${z}$, respectively. The $\zeta$ is a gauge variable and $g$ denotes the coupling constant. This Lagrangian respects 
the following gauge symmetry transformations $(\delta_{g})$:
\begin{equation}\label{gt}
\begin{split}
    &{\delta}_{g} r=0,\qquad {\delta}_{g}P_{r}=0,\qquad
    {\delta}_{g} \theta=g\lambda(t),\\ &{\delta}_{g} P_{\theta}=0,\qquad
    {\delta}_{g} z=\lambda(t),\qquad {\delta}_{g} P_{z}=0, \qquad {\delta}_{g}\zeta=\dot{\lambda}(t),
    \end{split}
\end{equation}
where $\lambda$ is the time dependent gauge parameter. The BRST invariant first-order Lagrangian $(L_{b})$ can be constructed by adding a BRST invariant function (i.e. gauge-fixing and Faddeev Popov ghost terms) into the first-order Lagrangian $L_{f}$ in \eqref{Lf} (cf. \cite{15} for details), as
\begin{eqnarray}\label{ab}
{L}_{b} &=& P_{r}\,\dot{r} + P_{\theta}\,\dot{\theta} + P_{z}\,\dot{z} - \frac{1}{2}\,p_{r}^{2} - \frac{1}{2r^{2}}\,p_{\theta}^{2} - \frac{1}{2}\,P_{z}^{2} - \zeta \big(P_z + g\,P_\theta \big)\nonumber\\
&-& V(r) + b \Big(\dot{\zeta} - z - \frac{1}{2}\,b\Big) + {\dot{\bar{\cal{C}}}}\, \dot{\cal{C}} + \bar{\cal{C}} \,{\cal{C}}.
\end{eqnarray}
Here, $(\bar{\cal{C}}){\cal{C}}$ are the Faddeev-Popov (anti-)ghost variables and $b$ is the Nakanishi-Lautrup auxiliary variable. The off-shell nilpotent and absolutely anti-commuting (anti-)BRST symmetry transformations $(s_{(a)b})$ can be given as
\begin{equation}\label{BRST}
\begin{split}
    &s_{b} r=0,\qquad s_{b} P_{r}=0,\qquad
    s_{b} \theta=g\,{\cal{C}},\qquad s_{b} P_{\theta}=0,\qquad s_{b} z={\cal{C}},\\&
     s_{b} P_{z}=0, \qquad s_{b}\zeta=\dot{\cal{C}},\qquad
   s_{b}{\cal{C}}=0,\qquad s_{b}{\cal\bar{C}}=b,\qquad s_{b} b=0,
    \end{split}
\end{equation}
and
\begin{equation}\label{ABRST}
\begin{split}
    &s_{ab} r=0,\qquad s_{ab} P_{r}=0,\qquad
    s_{ab} \theta=g\,{\cal\bar{C}},\qquad s_{ab} P_{\theta}=0,\qquad s_{ab} z={\cal\bar{C}},\\&
     s_{ab} P_{z}=0, \qquad s_{ab}\zeta={\dot{\bar{\cal{C}}}},\qquad
   s_{ab}{\cal\bar{C}}=0,\qquad s_{ab}{\cal{C}}=-b,\qquad s_{ab} b=0.
    \end{split}
\end{equation}
The Lagrangian $(L_{b})$ in~\eqref{ab} remains quasi-invariant under the above set of (anti-)BRST symmetry transformations as one can check
\begin{equation}
s_{b}L_{b}=\frac{d}{dt} \big(b\,\dot{\cal{C}} \big), \qquad 
s_{ab}L_{b}=\frac{d}{dt} \big(b \,\dot{\bar{\cal{C}}}\big).
\end{equation}
According to the Noether's theorem, these continuous symmetries lead to the following conserved (anti-)BRST charges $(Q_{(a)b})$
\begin{eqnarray}
Q_{b}&=& b\,\dot{\cal C} + \big(P_z+ g\,P_{\theta} \big)\,{\cal{C}}  \equiv b\,\dot {\cal C} - {\dot b}\, {\cal C}, \nonumber\\
Q_{ab}&=&  b\,{\dot{\bar{\cal C}}} + \big(P_z + g\,P_{\theta} \big)\,\bar{\cal{C}} \equiv b\,{\dot {\bar {\cal C}}} - {\dot b}\, {\bar {\cal C}},
\end{eqnarray}
which are the generators of the (anti-)BRST transformations. These conserved charges are nilpotent of order two $(\text{i.e.} \; Q_{(a)b}^{2}=0)$ and absolutely anti-commuting $(\text{i.e.} \; Q_{b}\,Q_{ab}+Q_{ab}\,Q_{b}=0)$ in nature. The Lagrangian $(L_{b})$ in~\eqref{ab} also respects other nilpotent and anti-commuting continuous symmetries known as (anti-)co-BRST symmetries or (anti-)dual-BRST symmetries. The following (anti-)co-BRST symmetry transformations $(s_{(a)d})$
\begin{equation}\label{coBRST}
\begin{split}
    &s_{d}r=0,\qquad s_{d} P_{r}=0,\qquad
    s_{d} \theta=g\,{\dot{\bar{\cal{C}}}},\qquad s_{d} P_{\theta}=0,\qquad s_{d} z={\dot{\bar{\cal C}}},\\&
     s_{d} P_{z}=0, \qquad s_{d}\zeta={\bar{\cal C}},\qquad
   s_{d}{\bar{\cal{C}}}=0,\qquad s_{d}{\cal C}=- \big(P_{z}+g\,P_{\theta} \big),\qquad s_{d} b=0,
    \end{split}
\end{equation}
\begin{equation}\label{acoBRST}
\begin{split}
    &s_{ad} r=0,\qquad s_{ad} P_{r}=0,\qquad
    s_{ad} \theta=g\,{\cal\dot{C}},\qquad s_{ad} P_{\theta}=0,\qquad s_{ad} z={\cal\dot{C}},\\&
     s_{ad} P_{z}=0, \qquad s_{ad}\zeta={\cal{C}},\qquad
   s_{ad}{\cal\bar{C}}=\big(P_{z}+g\,P_{\theta}\big),\qquad s_{ad}{\cal{C}}=0,\qquad s_{ad} b=0,
    \end{split}
\end{equation}
leave the Lagrangian $(L_{b})$ quasi-invariant as:
\begin{eqnarray}\label{z}
s_{d}L_{b}=\dfrac{d}{dt}\Big(\big(P_z + g\,P_\theta \big) {\dot{\bar{\cal C}}} \Big), \qquad 
s_{ad}L_{b} =\dfrac{d}{dt} \Big(\big(P_z + g\,P_\theta \big)\dot{\cal{C}} \Big).
\end{eqnarray}
The conserved charges corresponding to the (anti-)co-BRST symmetry transformations are obtained as follows:
\begin{eqnarray}
Q_d &=& b\,{\bar{\cal{C}}} + \big(g\,P_{\theta} + P_{z} \big){\dot{\bar{\cal{C}}}}  \equiv b \, {\bar {\cal C}} - \dot b\,{\dot {\bar {\cal C}}}, \nonumber\\ 
Q_{ad} &=&  b\, {\cal C} + \big(g\,P_{\theta} + P_{z} \big){\dot{\cal{C}}}  \equiv b \, {\cal C} - \dot b\,{\dot {\cal C}}.
\end{eqnarray}
These nilpotent and anti-commuting charges are the generators of the (anti-)co-BRST symmetry transformations. The (anti-)BRST and (anti-)co-BRST transformations differ primarily in a sense that (anti-)BRST transformations preserve the invariance of kinetic terms, whereas (anti-)co-BRST transformations leave the gauge-fixing term invariant.

\section{(Anti-)BRST Symmetries: Supervariable Approach}\label{S3}
In this section, we investigate the FLPR model within the framework of augmented supervariable approach, in which we utilize the horizontality condition together with the gauge invariant restrictions, to procure the complete set of (anti-)BRST symmetries of the system.
We begin with defining the exterior derivative $(d)$ and a one-form connection $(\zeta^{(1)})$ on $(0+1)$-dimensional ordinary space parametrized by time evolution parameter $t$ as follows:
\begin{equation}
    d=dt\,\partial_{t}, \qquad \zeta^{(1)}=dt\,\zeta(t).
\end{equation}
Thus, the action of exterior derivative on a one-form yields two-form
\begin{equation}
d\,\zeta^{(1)} = 0, \qquad\text{since}\; \big(dt\wedge dt \big)=0.
\end{equation}
Now, to make use of supervariable approach, we generalize the exterior derivative $(d)$ and one-form $(\zeta^{(1)})$ to their corresponding counterpart super-exterior derivative $(\widetilde{d})$ and super-one-form $(\widetilde{\zeta}^{(1)})$,  respectively on the $(1,2)$-dimensional superspace parameterized by time evolution parameter $t$ and a pair of Grassmannian variables $(\eta$ and $\bar{\eta})$. Thus, on  $(1,2)$-dimensional superspace, we have the following generalizations, namely;
\begin{equation}
\begin{split}
    &~~~\widetilde{d} = dt\,\partial_{t}+d\eta\,\partial_{\eta}+d\bar{\eta}\,\partial_{\bar{\eta}},\\
    &\widetilde{{\zeta}}^{(1)} = dt\,{\Xi}(t,\eta,\bar{\eta})+d\eta\,{\cal{\bar{F}}}(t,\eta,\bar{\eta})+d\bar{\eta}\,{\cal{F}}(t,\eta,\bar{\eta}),
    \end{split}
\end{equation}
where $\partial_{\eta}=\frac{\partial}{\partial\eta}$ and $\partial_{\bar{\eta}}=\frac{\partial}{\partial\bar{\eta}}$ are the Grassmannian derivatives, which are nilpotent of order two $({\text i.e.} \; \partial_\eta^{2} = \partial_{\bar{\eta}}^{2}=0)$ and absolutely anti-commuting $({\text i.e.} \; \partial_\eta \partial_{\bar{\eta}} + \partial_{\bar{\eta}} \partial \eta = 0)$. Expanding the components of super-one-form (or super multiplets) along the Grassmannian directions  $\eta$ and $\bar{\eta}$ as
\begin{equation}\label{sm}
\begin{split}
&{\Xi}(t,\eta,\bar{\eta})=\zeta(t)+\eta\,\bar{f_{1}}(t)+\bar{\eta}\,f_{1}(t)+\eta\bar{\eta}\,B(t),\\
&{\cal{F}}(t,\eta,\bar{\eta})={\cal{C}}(t)+\eta\,{\bar b}_1(t)+\bar{\eta}\,b_{1}(t)+\eta\bar{\eta}\,s(t),\\
&{\cal{\bar{F}}}(t,\eta,\bar{\eta})={\cal{\bar{C}}}(t)+\eta\,{\bar b}_2(t)+\bar{\eta}\,b_{2}(t)+\eta\bar{\eta}\,\bar{s}(t),
\end{split}
\end{equation}
where $f_{1}$, $\bar{f}_{1}$, $s$ and $\bar{s}$ are the fermionic secondary variables and $B$, $b_{1}$, ${\bar b}_1$, $b_{2}$ and ${\bar b}_2$ are bosonic secondary variables. Now, to express all supervariables in terms of the basic and auxiliary variables, we enforce the so called horizontality condition. The horizontality condition demands that the curvature two-form should be independent of the Grassmannian variables $\eta$ and $\bar{\eta}$. i.e.,
\begin{equation}\label{hc}
    \widetilde{d}\,\widetilde{\zeta}^{(1)}=d\,\zeta^{(1)}.
\end{equation}
Thus, the two-form in (1,2)-dimensional superspace can be written as
\begin{eqnarray}
\widetilde{d}\,\widetilde{\zeta}^{(1)}&=&(dt\wedge d\eta)\big(\partial_{t}\bar{\cal{F}}-\partial_{\eta}\Xi \big)+(dt\wedge d\bar{\eta})\big(\partial_{t}{\cal{F}}-\partial_{\bar{\eta}}\Xi \big)\nonumber\\
&-& (d\eta\wedge d\bar{\eta})\big(\partial_{\eta}{\cal{F}}+\partial_{\bar{\eta}}\bar{\cal{F}} \big)-(d\eta\wedge d\eta)\partial_{\eta}\bar{\cal{F}}- (d\bar{\eta}\wedge d\bar{\eta})\partial_{\bar{\eta}}{\cal{F}}.
\end{eqnarray}
Using the horizontality condition~\eqref{hc} and expansion of supermultiplets~\eqref{sm}, we obtain  following expression for secondary variables in terms of the basic dynamical and auxiliary variables as
\begin{equation}
\begin{split}
    &\bar{f}_{1}={\dot{\bar{\cal{C}}}},\qquad f_{1}=\dot{\cal{C}},\qquad b_{1}=0,\qquad b_2 + {\bar b}_1=0,\\ &\dot{b}_{2}=B,\qquad \bar{b}_{2}=0,\qquad s=0,\qquad \bar{s}=0.
    \end{split}
\end{equation}
Now, we make a choice  $b_{2}=-{\bar b}_1 = b$ for the Nakanishi-Lautrup type auxiliary variable and substituting the above relations in \eqref{sm} leads to the following desired expressions for the supervariables 
\begin{equation}\label{sm1}
    \begin{split}
        &{\Xi}^{(h)}(t,\eta,\bar{\eta})=\zeta(t)+\eta\,{\dot{\bar{\cal{C}}}}+\bar{\eta}\,\dot{\cal{C}}+\eta{\bar{\eta}}\,{\dot{b}},\\
        &{\cal{F}}^{(h)}(t,\eta,\bar{\eta})={\cal{C}}(t)-\eta \,b,\\
        &{\cal{\bar{F}}}^{(h)}(t,\eta,\bar{\eta})={\cal{\bar{C}}}(t)+\bar{\eta} \,b,
    \end{split}
\end{equation}
where the superscript $(h)$ indicates that the superexpansions are being obtained after utilizing the horizontality condition. So far, we have attained the complete form of supervariables to procure the (anti-)BRST transformations for the variables $\zeta, {\cal{C}}$ and $\bar{\cal{C}}$.

To procure the (anti-)BRST symmetry transformations for the dynamical variable $z$ and $\theta$, we exploit the gauge invariant restrictions. Specifically, the gauge invariant quantities would be independent of the Grassmannian variables when generalized on superspace. To this end, the quantities $(\zeta-\dot{z})$ and $(g\,\zeta-\dot{\theta})$ remain invariant under the gauge transformations (cf. \eqref{gt}) and it will accomplish our purpose of getting the off-shell nilpotent (anti-)BRST symmetry transformations of the variables $z$ and $\theta$, respectively. Thus, we express these quantities in the language of differential forms as
\begin{eqnarray}
    &\zeta^{(1)}-dz^{(0)}=dt\;\big(\zeta(t)-\partial_{t}z(t)\big),\nonumber\\
    &g\,\zeta^{(1)}-d\theta^{(0)}=dt\;\big(g\zeta(t)-\partial_{t}\theta(t)\big),
\end{eqnarray}
where $\zeta^{(1)}$ is a one-form and $z^{(0)}$, $\theta^{(0)}$ define the zero-forms. Generalizing these one-form quantities onto (1,2)-dimensional superspace in the following fashion:
\begin{eqnarray}\label{1}
    &\widetilde{\zeta}^{(1)}-\widetilde{d}\,\widetilde{z}^{(0)}=\zeta^{(1)}-dz^{(0)},\nonumber\\
    &g\,\widetilde{\zeta}^{(1)}-\widetilde{d}\,\widetilde{\theta}^{(0)}=g\zeta^{(1)}-d\theta^{(0)}.
\end{eqnarray}
These super-zero-forms can be expanded along the Grassmannian directions as
\begin{eqnarray}\label{2}
&\widetilde{z}^{(0)}={\cal{Z}}(t,\eta,{\bar{\eta}})=z(t) + \eta\,\bar{f}_{2}(t) + \bar{\eta}f_{2}(t) + \eta\bar{\eta}\,\bar{B}(t),\nonumber\\
&\widetilde{\theta}^{(0)}=\Theta(t,\eta,\bar{\eta})=\theta(t)+\eta\,\bar{f}_{3}(t) + \bar{\eta}\,f_{3}(t) + \eta\bar{\eta}\,b_{3}(t).
\end{eqnarray}
Here $f_{2}$, $\bar{f_{2}}$, $f_{3}$, $\bar{f_{3}}$ are fermionic secondary variables and $\bar{B}$, $b_{3}$ are bosonic secondary variables. The gauge [or (anti-)BRST] invariant quantities in~\eqref{1} can be explicitly expressed as
\begin{eqnarray}\label{3}
    &\widetilde{\zeta}^{(1)}-\widetilde{d}\,\widetilde{z}^{(0)}=dt\,\big(\Xi^{(h)}-\partial_{t}{\cal{Z}} \big)+d\eta\,\big(\bar{\cal{F}}^{(h)}-\partial_{\eta}{\cal{Z}}\big) + d\bar{\eta} \big({\cal{F}}^{(h)}-\partial_{\bar{\eta}}{\cal{Z}}\big),\nonumber\\
    &g\,\widetilde{\zeta}^{(1)}-\widetilde{d}\,\widetilde{\theta}^{(0)}=dt\,\big(g\,\Xi^{(h)}-\partial_{t}\Theta \big) + d\eta\,\big(g\,\bar{\cal{F}}^{(h)}-\partial_{\eta}\Theta \big) + d\bar{\eta}\,\big(g\,{\cal{F}}^{(h)}-\partial_{\bar{\eta}}\Theta\big).
\end{eqnarray}
Using Eqs.~\eqref{sm1}--\eqref{3}, we deduce the values of all secondary variables as follows:
\begin{eqnarray}
    &f_{2}={\cal{C}},\qquad \bar{f_{2}}=\bar{{\cal{C}}},\qquad \bar{B}=b,\nonumber\\
    &f_{3}=g\,{\cal{C}},\qquad \bar{f}_{3}=g\,\bar{{\cal{C}}},\qquad b_{3}=g\,b.
    \end{eqnarray}
Thus, the super expansion for the supervaribles ${\cal{Z}}$ and $\Theta$, can be respectively given as
\begin{eqnarray}\label{sv1}
&{\cal{Z}}^{(h)}(t,\eta,{\bar{\eta}})=z(t)+\eta\,\bar{{\cal{C}}}(t) + \bar{\eta}\,{\cal{C}}(t)+\eta\bar{\eta}\,b(t),\nonumber\\
&\Theta^{(h)}(t,\eta,\bar{\eta}) = \theta(t) + g\eta\,\bar{{\cal{C}}}(t) + g\bar{\eta}\,{\cal{C}}(t) + g\eta\bar{\eta}\,b(t).
\end{eqnarray}
Moreover, the dynamical variables $r, P_{r},P_{\theta}\;\text{and}\;P_{z}$ are gauge invariant. Therefore, these variables remain unaffected by the presence of the Grassmannian variables in superspace, i.e.;
\begin{eqnarray}\label{sv2}
    &{\cal{R}}^{(h)}(t,\eta,{\bar{\eta}})=r(t), \quad 
    &{\cal{P}}_{r}^{(h)}(t,\eta,{\bar{\eta}})=P_{r}(t),\nonumber\\
    &{\cal{P}}_{\theta}^{(h)}(t,\eta,{\bar{\eta}})=P_{\theta}(t), \quad 
    &{\cal{P}}_{z}^{(h)}(t,\eta,{\bar{\eta}})=P_{z}(t).
\end{eqnarray}
In view of the above super-expansions, we can deduce the nilpotent and absolutely anti-commuting (anti-)BRST symmetry transformations for any generic variables $(\phi)$ in the theory from its corresponding supervariables $\Phi^{(h)}(t,\eta,\bar{\eta})$ as
\begin{eqnarray}
    &s_{b}\phi(t)=\dfrac{\partial}{\partial\bar{\eta}}\Phi^{(h)}(t,\eta,{\bar{\eta}})\bigg|_{\eta=0},\qquad s_{ab}\phi(t)=\dfrac{\partial}{\partial{\eta}}\Phi^{(h)}(t,\eta,{\bar{\eta}})\bigg|_{\bar{\eta}=0},\nonumber\\&s_{b}s_{ab}\phi(t)=\dfrac{\partial}{\partial\bar{\eta}}\dfrac{\partial}{\partial{\eta}}\Phi^{(h)}(t,\eta,{\bar{\eta}}).
\end{eqnarray}
These expressions elucidate that the (anti-)BRST symmetry transformation of any generic variable is equivalent to translation of corresponding supervariable in the direction of Grassmannian variable $(\eta)\bar{\eta}$ on the $(1,2)$-dimensional superspace while keeping other direction fixed. Consequently, we derived the (anti-)BRST transformations for all the variables as listed in Eqs.~\eqref{BRST} and \eqref{ABRST}, except for the Nakanishi-Lautrup variable $b$, via the supervariable approach. The transformations of latter variable can be obtained by the requirements of anti-commutativity and nilpotency of the (anti-)BRST transformations.

\section{(Anti-)co-BRST Symmetries: Supervariable Approach}\label{S4}
In this section, we derive the off-shell nilpotent and absolutely anti-commuting (anti-)co-BRST symmetries of the FLPR model by making use of the dual-horizontality condition together with the (anti-)co-BRST invariant restrictions. We notice that the total gauge-fixing terms i.e. $b\big(\dot{\zeta}-z\big) -\frac{b^2}{2} = \frac{1}{2}(\dot \zeta - z)^2$ remain invariant under the (anti-)co-BRST transformations (cf. Eqs.\eqref{coBRST} and \eqref{acoBRST}). In terms of differential forms, the term $(\dot \zeta - z)$ can be written as
\begin{equation}
    \delta\zeta^{(1)}-z^{(0)}=\dot{\zeta}(t) - z(t),
\end{equation}
where $\delta=*\,d\,*$ is the co-exterior derivative and $*$ is the Hodge duality operation defined on one-dimensional ordinary space. Thus, we can say that the gauge-fixing term has its origin in the co-exterior derivative. The co-exterior derivative acting on a one-form $\zeta^{(1)}$ results in
\begin{equation}
\delta\zeta^{(1)}=\big(*\,d\,*\big)\zeta^{(1)}=\dot \zeta(t).
\end{equation}
In order to derive the complete set of (anti-)co-BRST symmetry transformations, we generalize the co-exterior derivative $(*\;d\;*)$ and one-form $(\xi^{(1)})$ to corresponding super-co-exterior derivative $(\star\;\Tilde{d}\;\star)$ and super-one-form $(\Tilde{\xi}^{(1)})$ on the $(1,2)$-dimensional superspace.  Consequently, the dual-horizontality condition implies that 
\begin{equation}
   \big(\star\,\widetilde{d}\,\star \big) \widetilde{\zeta}^{(1)} - \widetilde z^{(0)}= \big(*\,d\,*\big)\zeta^{(1)}-z^{(0)},
\end{equation}
where $\star$ is the Hodge duality operation  defined  on $(1, 2)$-dimensional supermanifold. Now expanding in terms of supervariables, we obtain
\begin{equation}\label{dhc}
\big(\partial_{t}\Xi+\partial_{\eta}\bar{{\cal{F}}}
+ \partial_{\bar{\eta}}{\cal{F}} \big) 
+ S^{\eta\eta} \big(\partial_{\eta}{\cal{F}} \big) 
+ S^{\bar{\eta}\bar{\eta}} \big(\partial_{\bar{\eta}}\bar{{\cal{F}}} \big) - {\cal{Z}}=\dot{\zeta}-z(t),
\end{equation}
where $S^{\eta\eta}$, $S^{\bar{\eta}\bar{\eta}}$ and $S^{\eta \bar\eta}$ are symmetric in $\eta$ and $\bar{\eta}$. Here we used the working principles for the Hodge duality $\star$ operation on a $(1,2)$-dimensional superspace~\cite{8,18};
\begin{eqnarray}\label{hdw}
&&\star\,dt=(d\eta\wedge d\bar{\eta}),\qquad \star\,(dt\wedge d\eta\wedge d\bar{\eta})= S^{\eta\bar\eta} =1, \qquad (dt\wedge dt\wedge d\eta)=0, \nonumber\\
&& \star\, d\eta=(dt\wedge d\bar{\eta}),\qquad\star\,(dt\wedge d\eta\wedge d\eta)=S^{\eta\eta},  \qquad \qquad (d\eta\wedge d\bar{\eta}\wedge d\bar{\eta})=0, \nonumber\\
&& \star\, d\bar{\eta}=(dt\wedge d{\eta}),\qquad
\star\,(dt\wedge d\bar{\eta}\wedge d\bar{\eta})=S^{\bar{\eta}\bar{\eta}},\qquad \qquad
(d\eta\wedge d\eta\wedge d\bar{\eta})=0.
\end{eqnarray}
Exploiting the Eqs. \eqref{sm} and \eqref{dhc} yields the following relationships:
\begin{eqnarray}
&&\dot{\bar{f}}_{1}=\bar{f}_{2},\qquad \dot{f}_{1}=f_{2},\qquad \dot{B}=\bar{B},\qquad b_{1}=-\bar{b}_{2},\nonumber\\
&&\bar{b}_{1}=0,\qquad \;\; b_{2}=0,\qquad \;\; s=0,\qquad \;\;\bar{s}=0.
\end{eqnarray}
Substituting these relations in Eqs.~\eqref{sm} and \eqref{2} and making the choice $b_{1}=-\bar{b}_{2}={\cal{B}}$, we obtain the super-expansion for the supervariables as
\begin{eqnarray}\label{rsv}
    {\cal{Z}}^{(r)}(t,\eta,\bar{\eta})&=&z(t)+\eta\,\dot{\bar{f}}_{1}+\bar{\eta}\,\dot{f_{1}}+\eta\bar{\eta}\,\dot{B},\nonumber\\
    {\cal{F}}^{(r)}(t,\eta,\bar{\eta})&=&{\cal{C}}(t)+\bar{\eta}\,{\cal{B}},\nonumber\\
\bar{{\cal{F}}}^{(r)}(t,\eta,\bar{\eta})&=&\bar{\cal{C}}(t)-\eta\, {\cal{B}},
\end{eqnarray}
where superscript $(r)$ denotes the reduced form of the supervariables. Nevertheless, we are unable to derive the super-expansion of the supervariable $\Theta(t,\eta,\bar{\eta})$. Accordingly, we exploit (anti-)co-BRST invariant term $(g\,\dot{\zeta}-\theta)$, which would be independent of the Grassmannian variables upon the generalization on superspace. Explicitly, it can be written as
\begin{equation}
g\big(\star\,\widetilde{d}\,\star\big)\,\widetilde{\zeta}^{(1)}-\widetilde \theta^{(0)} = g\big(*\,d\,*\big)\zeta^{(1)}-\theta^{(0)}.
\end{equation}
Expanding using co-exterior derivative and supervariables, we have
\begin{equation}\label{dhc1}
    g\big(\partial_{t}\Xi+\partial_{\eta}\bar{{\cal{F}}}
    +\partial_{\bar{\eta}}{\cal{F}} \big)+g\,S^{\eta\eta}\big(\partial_{\eta}{\cal{F}}\big)
    +g\,S^{\bar{\eta}\bar{\eta}} \big(\partial_{\bar{\eta}}\bar{{\cal{F}}} \big)-\Theta=g\,\dot{\zeta}-\theta(t),
\end{equation}
where we have again used Eq.~\eqref{hdw}. Further, we obtain the relationships connecting secondary variables and generic variables as follows:
\begin{eqnarray}
&&g\dot{\bar{f}}_{1}=\bar{f}_{3},\qquad g\dot{f}_{1}=f_{3},\qquad g\dot{B}=b_3,\qquad b_{1}=-\bar{b}_{2},\nonumber\\
&&\bar{b}_{1}=0,\qquad \quad b_{2}=0,\qquad\quad  s=0,\qquad \quad\; \bar{s}=0.
\end{eqnarray} 
Thus, we have derived the reduced form of supervariable $\Theta(t,\eta,\bar{\eta})$ as
\begin{equation}
\Theta^{(r)}(t,\eta,\bar{\eta})=\theta(t)+g\,\eta\,\dot{\bar{f}}_{1}+g\,\bar{\eta}\,\dot{f_{1}}+g\,\eta\bar{\eta}\,\dot{B}.
\end{equation}
However, it is evident that we have not yet obtained the cherished super-expansion for the supervariables in terms of basic and auxiliary variables present in the theory. To obtain the exact super-expansion, we use the following (anti-)co-BRST invariant restrictions on the dynamical variables;
\begin{eqnarray}
&& s_{(a)d}\big[\zeta \big(P_{z}+g\,P_{\theta} \big)-\bar{\cal{C}}\,{\cal{C}} \big]=0. 
\end{eqnarray}
We generalize the above (anti-)co-BRST invariant restriction on to $(1,2)$-dimensional supermanifold and demand that these restrictions must be independent of the Grassmannian variables $\eta$ and $\bar{\eta}$. Mathematically, they can be expressed as follows:
\begin{eqnarray}\label{4}
    \Xi\big({\cal{P}}_{z}+g\,{\cal{P}}_{\theta}\big)-\bar{\cal{F}}^{(r)}{\cal{F}}^{(r)}=\zeta\big(P_{z}+g\,P_{\theta}\big)-\bar{\cal{C}}\,{\cal{C}}.
\end{eqnarray}
It should be noted that the dynamical variables $r$, $P_r$, $P_z$ and $P_\theta$ remain invariant under the (anti-)co-BRST symmetry transformations [i.e. $s_{(a)d} (r, \,P_r,\, P_z, \, P_\theta) = 0$]. Therefore, these variables also remain unaffected when generalizing them on $(1, 2)$-dimensional supermanifold, i.e.;
\begin{eqnarray}\label{41}
{\cal R}(t, \eta,\bar\eta) = r(t), \qquad
{\cal P}_r(t, \eta,\bar\eta) = P_r(t), \nonumber\\
{\cal P}_z(t, \eta,\bar\eta) = P_z(t), \qquad
{\cal P}_\theta(t, \eta,\bar\eta) = P_\theta(t).
\end{eqnarray}
Substituting the above super-expansion  of supervariables listed in Eqs.~(\ref{rsv}), (\ref{41}) into Eq.~(\ref{4}) and equating various coefficients, we obtain the following relations
\begin{eqnarray}
    \bar{f}_{1} \big(P_{z} + g\,P_{\theta}\big)+{\cal{B}}\,{\cal{C}}=0,\quad f_{1}\big(P_{z}+g\,P_{\theta}\big)+{\cal{B}}\,\bar{\cal{C}}=0,\quad B\big(P_{z}+g\,P_{\theta}\big)+{\cal{B}}\,{\cal{B}}=0.
\end{eqnarray}
We notice that this is not sufficient to determine the secondary variables in terms of dynamic and auxiliary variables. Therefore, we invoke co-BRST and anti-co-BRST invariant restrictions: $s_{d}(\zeta\,\bar{\cal{C}})=0$ and $s_{ad}(\zeta\,{\cal{C}})=0$, respectively which are explicitly written in $(1,2)$-dimensional superspace as
\begin{eqnarray}
   \Xi\,\bar{\cal{F}}^{(r)}=\zeta\,\bar{\cal{C}},\qquad \Xi\,{\cal{F}}^{(r)}=\zeta\,{\cal{C}}.
\end{eqnarray}
This leads to the following set of expressions:
\begin{eqnarray}
   &&f_{1}\,\bar{\cal{C}}=0,\hspace{3.1cm}\bar{f_{1}}\,{\cal{C}}=0,\nonumber\\
   &&\bar{f}_{1}\,\bar{\cal{C}}-\zeta\,{\cal{B}}=0,\hspace{2cm}f_{1}\,{\cal{C}}+{\cal{B}}\,\zeta=0,\nonumber\\
  && B\,\bar{\cal{C}}-f_{1}\,{\cal{B}}=0,\hspace{2cm}B\,{\cal{C}}-\bar{f}_{1}\,{\cal{B}}=0.
    \end{eqnarray}
The solutions of these equations can be written as
\begin{equation}
    f_{1}=\bar{\cal{C}},\qquad \bar{f}_{1}={\cal{C}},\qquad {\cal{B}}=B=-\big(P_{z}+gP_{\theta}\big).
\end{equation}
Finally, we can write the super-expansion of all supervariables in the theory. 
Thus, we obtain the complete set of super-expansions along the direction of the Grassmannian variables as:
\begin{eqnarray}\label{sv3}
    {\cal{Z}}^{(d)}(t,\eta,\bar{\eta})&=&z(t)+\eta\,\dot{\cal{C}}+{\bar{\eta}}\,{\dot{\bar{\cal{C}}}}-\eta\bar{\eta}\,\big(\dot{P}_{z}+g\,\dot{P_{\theta}}\big),\nonumber\\
    \Theta^{(d)}(t,\eta,\bar{\eta})&=&\theta(t)+ g \,\eta\,\dot{\cal{C}} + g\,{\bar{\eta}}\,{\dot{\bar{\cal{C}}}}-g\,\eta\bar{\eta}\,\big(\dot{P}_{z} + g\,\dot{P_{\theta}}\big),\nonumber\\
    \Xi^{(d)}(t,\eta,\bar{\eta})&=&\zeta(t)+\eta\,{\cal{C}} + {\bar{\eta}}\,{{\bar{\cal{C}}}}-\eta\bar{\eta}\,\big({P}_{z}+g\,{P_{\theta}} \big),\nonumber\\
    {\cal{F}}^{(d)}(t,\eta,\bar{\eta})&=&{\cal{C}}(t)-\bar{\eta}\big({P}_{z}+g\,{P_{\theta}}\big),\nonumber\\
    \bar{\cal{F}}^{(d)}(t,\eta,\bar{\eta})&=&\bar{\cal{C}}(t)+{\eta}\big({P}_{z} + g\,{P_{\theta}}\big),\nonumber\\
    {\cal{R}}^{(d)}(t,\eta,\bar{\eta})&=&r(t), \qquad 
    {\cal{P}}_{r}^{(d)}(t,\eta,\bar{\eta}) = P_{r}(t),\nonumber\\
    {\cal{P}}_{\theta}^{(d)}(t,\eta,\bar{\eta})&=&P_{\theta}(t), ~\quad
    {\cal{P}}_{z}^{(d)}(t,\eta,\bar{\eta}) =P_{z}(t),
\end{eqnarray}
where the superscript $(d)$ denotes that the super-expansions are obtained after utilizing the dual-horizontality condition. We can now procure the complete set of off-shell nilpotent and absolutely anti-commuting (anti-)co-BRST symmetries of the system
\begin{eqnarray}
     &s_{d}\phi(t)=\dfrac{\partial}{\partial\bar{\eta}}\Phi^{(d)}(t,\eta,{\bar{\eta}})\bigg|_{\eta=0},\qquad s_{ad}\phi(t)=\dfrac{\partial}{\partial{\eta}}\Phi^{(d)}(t,\eta,{\bar{\eta}})\bigg|_{\bar{\eta}=0},\nonumber\\&s_{d}s_{ad}\phi(t)=\dfrac{\partial}{\partial\bar{\eta}}\dfrac{\partial}{\partial{\eta}}\Phi^{(d)}(t,\eta,{\bar{\eta}}). 
\end{eqnarray}
Essentially, the (anti-)co-BRST transformation of the generic variable is equivalent to the translation of the corresponding supervariable along the Grassmannian direction $(\eta)\bar{\eta}$ while keeping the other direction $(\bar{\eta})\eta$ fixed.

\section{Key Features of (Anti-)BRST and (Anti-)co-BRST Symmetries: Supervariable Approach}\label{S5}
In this section, we encapsulate the key properties such as nilpotency and anti-commutativity of the (anti-)BRST and (anti-)co-BRST symmetry transformations in terms of supervariables and Grassmannian translational generators. Moreover, we capture the invariance of the Lagrangian under (anti-)BRST and (anti-)co-BRST symmetry transformations within the framework of the supervariable approach.

\subsection{Nilpotency and Anti-commutativity}
The Nilpotency of order two and absolute anti-commutativity are the two basic features of the (anti-)BRST and (anti-)co-BRST symmetry transformations. In a $(1,2)$-dimensional superspace, the nilpotency property for any generic variable can be expressed in the following manner
\begin{eqnarray}
    s_{b}^{2}\phi(t)=0\;\Longleftrightarrow\;\frac{\partial}{\partial\bar{\eta}}\frac{\partial}{\partial\bar{\eta}}\Phi^{(h)}(t,\eta,\bar{\eta})=0,\nonumber\\
    s_{ab}^{2}\phi(t)=0\;\Longleftrightarrow\;\frac{\partial}{\partial{\eta}}\frac{\partial}{\partial{\eta}}\Phi^{(h)}(t,\eta,\bar{\eta})=0,\nonumber\\
    s_{d}^{2}\phi(t)=0\;\Longleftrightarrow\;\frac{\partial}{\partial\bar{\eta}}\frac{\partial}{\partial\bar{\eta}}\Phi^{(d)}(t,\eta,\bar{\eta})=0,\nonumber\\
    s_{ad}^{2}\phi(t)=0\;\Longleftrightarrow\;\frac{\partial}{\partial{\eta}}\frac{\partial}{\partial{\eta}}\Phi^{(d)}(t,\eta,\bar{\eta})=0,
    \end{eqnarray}
where $\phi(t)$ is any generic variable, $\Phi^{(h)}(t,\eta,\bar{\eta})$ and $\Phi^{(d)}(t,\eta,\bar{\eta})$ are its corresponding supervariable [cf. Eqs.\eqref{sm1}, \eqref{sv1}, \eqref{sv2} and \eqref{sv3}]. In a similar fashion, the absolute anti-commutativity reads
\begin{eqnarray}
    \big(s_{b}s_{ab}+s_{ab}s_{b}\big)\phi(t)=0\;\Longleftrightarrow\; \bigg(\frac{\partial}{\partial\bar{\eta}}\frac{\partial}{\partial\eta}+\frac{\partial}{\partial\eta}\frac{\partial}{\partial\bar{\eta}} \bigg)
    \Phi^{h}(t,\eta,\bar{\eta})=0,\nonumber\\
    \big(s_{d}s_{ad}+s_{ad}s_{d}\big)\phi(t)=0\;\Longleftrightarrow\;\bigg(\frac{\partial}{\partial\bar{\eta}}\frac{\partial}{\partial\eta}+\frac{\partial}{\partial\eta}\frac{\partial}{\partial\bar{\eta}} \bigg)
    \Phi^{d}(t,\eta,\bar{\eta})=0.
\end{eqnarray}
In addition, we also explore the nilpotency and anti-commutativity of (anti-)BRST and (anti-)co-BRST charges and provide their geometrical interpretation in terms of Grassmannian translational generators. To begin with, the BRST and anti-BRST charges can be expressed in terms of BRST and/or anti-BRST transformations as
\begin{eqnarray}
    &&Q_{b}=s_{b}\big(\bar{\cal{C}}\,\dot{\cal{C}} - {\dot{\bar{\cal{C}}}}\,{\cal{C}} \big)=s_{ab} \big(\dot{\cal{C}}\,{\cal{C}} \big)=-s_{b}s_{ab}\big(\zeta\,{\cal{C}}\big)=-\frac{1}{2}\,s_{b}s_{ab}\big({\dot{z}}\,{\cal{C}}-z\,{\cal{\dot{C}}}\big),\nonumber\\
    &&Q_{ab}=-s_{ab}\big(\bar{\cal{C}}\,\dot{\cal{C}} - {\dot{\bar{\cal{C}}}}\, {\cal{C}}\big)=-s_{b}\big({\dot{\bar{\cal{C}}}}\,\bar{\cal{C}}\big)=-s_{b}s_{ab}\big(\zeta\,\bar{\cal{C}}\big)=-\frac{1}{2}\,s_{b}s_{ab}\big({\dot{z}}\,{\bar{\cal{C}}}-z\,{\cal{\dot{\bar{C}}}} \big).
\end{eqnarray}
To provide an analogy in $(1,2)$-dimensional superspace, we have
\begin{eqnarray}
    Q_{b}&=&\frac{\partial}{\partial\bar{\eta}}\Big(\bar{\cal{F}}^{(h)}{\dot{\cal{F}}}^{(h)}-{\dot{\bar{\cal{F}}}}^{(h)}{\cal{F}}^{(h)}\Big)\bigg|_{\eta=0}
    \equiv \int d\bar{\eta}\;\Big(\bar{\cal{F}}^{(h)}{\dot{\cal{F}}}^{(h)}-{\dot{\bar{\cal{F}}}}^{(h)}{\cal{F}}^{(h)}\Big)\bigg|_{\eta=0}\nonumber\\
    &=&\frac{\partial}{\partial\eta}\Big({\dot{\cal{F}}}^{(h)}\,{\cal{F}}^{(h)}\Big)\bigg|_{\bar\eta = 0} \equiv \int d\eta\,\Big({\dot{\cal{F}}}^{(h)}\,{\cal{F}}^{(h)}\Big) \bigg|_{\bar\eta = 0}\nonumber\\
    &=&-\frac{\partial}{\partial\bar{\eta}}\frac{\partial}{\partial\eta}\Big(\Xi^{(h)}\,{\cal{F}}^{(h)}\Big) \equiv -\int d\bar{\eta}\int d\eta\, \Big(\Xi^{(h)}\,{\cal{F}}^{(h)}\Big)\nonumber\\
&=& -\dfrac{1}{2}\,\dfrac{\partial}{\partial \bar \eta} \dfrac{\partial}{\partial \eta} \Big({\dot {\cal Z}}^{(h)} {\cal F}^{(h)} - {\cal Z}^{(h)} {\dot {\cal F}}^{(h)} \Big) \equiv -\dfrac{1}{2} \int d{\bar \eta} \int d\eta \Big({\dot {\cal Z}}^{(h)} {\cal F}^{(h)} - {\cal Z}^{(h)} {\dot {\cal F}}^{(h)} \Big), 
\end{eqnarray}
and
\begin{eqnarray}
Q_{ab}&=&-\frac{\partial}{\partial{\eta}}\Big(\bar{\cal{F}}^{(h)}{\dot{\cal{F}}}^{(h)}-{\dot{\bar{\cal{F}}}}^{(h)}{\cal{F}}^{(h)}\Big)\bigg|_{\bar{\eta}=0} 
\equiv -\int d{\eta}\;\Big(\bar{\cal{F}}^{(h)}{\dot{\cal{F}}}^{(h)}-{\dot{\bar{\cal{F}}}}^{(h)}{\cal{F}}^{(h)}\Big)\bigg|_{\bar\eta=0} \nonumber\\
&=&-\frac{\partial}{\partial\bar{\eta}}\Big({\dot{\bar{\cal{F}}}}^{(h)} \bar{{\cal{F}}}^{(h)} \Big) \bigg|_{\eta = 0} 
\equiv -\int d\bar{\eta}\,\Big({\dot{\bar{\cal{F}}}}^{(h)}\bar{\cal{F}}^{(h)} \Big) \bigg|_{\eta = 0}\nonumber\\
&=&-\frac{\partial}{\partial\bar{\eta}}\frac{\partial}{\partial\eta}\Big(\Xi^{(h)}\,\bar{\cal{F}}^{(h)}\Big) \equiv -\int d\bar{\eta}\int d\eta\, \Big(\Xi^{(h)}\,\bar{\cal{F}}^{(h)}\Big) \nonumber\\
&=& -\dfrac{1}{2}\,\dfrac{\partial}{\partial \bar \eta} \dfrac{\partial}{\partial \eta} \Big({\dot {\cal Z}}^{(h)} {\bar {\cal F}}^{(h)} - {\cal Z}^{(h)} {\dot {\bar {\cal F}}}^{(h)} \Big) \equiv -\dfrac{1}{2} \int d{\bar \eta} \int d\eta \Big({\dot {\cal Z}}^{(h)} {\bar {\cal F}}^{(h)} - {\cal Z}^{(h)} {\dot {\bar {\cal F}}}^{(h)} \Big). \qquad 
\end{eqnarray}
It is noteworthy that, we can directly prove from the above expressions that $s_{b}Q_{b}=0$ and $s_{ab}Q_{ab}=0$, which sequentially imply the nilpotency of (anti-)BRST charges $({\text i.e.} \; Q_{b}^{2}=Q_{ab}^{2}=0)$. In addition, the relations $s_{b}Q_{ab}=0$ and $s_{ab}Q_{b}=0$ lead to the anti-commutativity $({\text i.e.} \; Q_{b}Q_{ab}+Q_{ab}Q_{b}=0)$ of the charges. Consequently, we write the nilpotency and anti-commutativity of (anti-)BRST charges in terms of the Grassmannian translational generators in the $(1,2)$-dimensional superspace as
\begin{eqnarray}
&& \frac{\partial}{\partial\bar{\eta}}Q_{b} = s_b Q_b =0\;\Longleftrightarrow\;Q_{b}^{2}=0,\nonumber\\
&& \frac{\partial}{\partial{\eta}}Q_{ab} =s_{ab} Q_{ab} =0\;\Longleftrightarrow\;Q_{ab}^{2}=0,\nonumber\\
 &&\frac{\partial}{\partial\bar{\eta}}Q_{ab} = s_b Q_{ab} =0\;\Longleftrightarrow\;Q_{b}\,Q_{ab}+Q_{ab}\,Q_{b}=0, \nonumber\\
 && \frac{\partial}{\partial\eta}Q_{b}= s_{ab} Q_b =0  \;\Longleftrightarrow\;Q_{b}\,Q_{ab}+Q_{ab}\,Q_{b}=0.
\end{eqnarray}
In an exact similar fashion, we can demonstrate the nilpotency and absolute anti-commutativity of the (anti-)co-BRST charges. In a one-dimensional space, (anti)-co-BRST charges takes the following form
\begin{eqnarray}
    &&Q_{d}=-s_{d}\big(\bar{\cal{C}}\,\dot{\cal{C}}-{\dot{\bar{\cal{C}}}\,{\cal{C}}}\big)=-s_{ad}\big({\dot{\bar{\cal{C}}}}\,\bar{\cal{C}} \big)=s_{d}s_{ad}\big(z\,\bar{\cal{C}}\big)=-\frac{1}{2}\,s_{d}s_{ad}\big(\zeta\,\dot{\bar{\cal{C}}}-\dot{\zeta}\,\bar{\cal{C}}\big),\nonumber\\
    &&Q_{ad}=s_{ad}\big(\bar{\cal{C}}\,\dot{\cal{C}} - {\dot{\bar{\cal{C}}}\,{\cal{C}}}\big)=s_{d}\big({\dot{{\cal{C}}}}\,{\cal{C}}\big)=s_{d}s_{ad}\big(z\,{\cal{C}}\big)=-\frac{1}{2}\,s_{d}s_{ad}\big(\zeta\,\dot{{\cal{C}}}-\dot{\zeta}\,{\cal{C}}\big).
    \end{eqnarray}
These co-BRST and anti-co-BRST charges can be expressed in terms of the Grassmannian translational generators in the superspace as follows: 
\begin{eqnarray}
Q_{d}&=&-\frac{\partial}{\partial\bar{\eta}}\Big(\bar{\cal{F}}^{(d)}{\dot{\cal{F}}}^{(d)}-{\dot{\bar{\cal{F}}}}^{(d)}{\cal{F}}^{(d)}\Big)\bigg|_{\eta=0} 
\equiv -\int d\bar{\eta}\,\Big(\bar{\cal{F}}^{(d)}{\dot{\cal{F}}}^{(d)}-{\dot{\bar{\cal{F}}}}^{(d)}{\cal{F}}^{(d)}\Big)\bigg|_{\eta=0}\nonumber\\
&=&-\frac{\partial}{\partial\eta}\Big({\dot{\bar{\cal{F}}}}^{(d)}
\bar{\cal{F}}^{(d)}\Big) \bigg|_{\bar \eta = 0}
\equiv -\int d\eta\, \Big({\dot{\bar{\cal{F}}}}^{(d)}\bar{\cal{F}}^{(d)}\Big)\bigg|_{\bar \eta = 0}\nonumber\\
&=&\frac{\partial}{\partial\bar{\eta}}\frac{\partial}{\partial\eta}\Big({\cal{Z}}^{(d)}\,\bar{\cal{F}}^{(d)}\Big) \equiv \int d\bar{\eta}\int d\eta\, \Big({\cal{Z}}^{(d)}\,\bar{\cal{F}}^{(d)}\Big)\nonumber\\
    &=&-\frac{1}{2}\,\frac{\partial}{\partial\bar{\eta}}\frac{\partial}{\partial\eta}\Big({\Xi}^{(d)}\,{\dot{\bar{\cal{F}}}}^{(d)}-{\dot{\Xi}}^{(d)}\,\bar{\cal{F}}^{(d)}\Big) \equiv -\frac{1}{2}\int d\bar{\eta}\int d\eta \,\Big({\Xi}^{(d)}\,{\dot{\bar{\cal{F}}}}^{(d)} - {\dot{\Xi}}^{(d)}\, \bar{\cal{F}}^{(d)}\Big), \qquad
\end{eqnarray}
and
\begin{eqnarray}
Q_{ad}&=&\frac{\partial}{\partial{\eta}}\Big(\bar{\cal F}^{(d)}{\dot{\cal F}}^{(d)}-{\dot {\bar {\cal F}}}^{(d)}{\cal{F}}^{(d)}\Big)\bigg|_{\bar{\eta}=0} \equiv
\int d{\eta}\;\Big(\bar{\cal{F}}^{(d)}{\dot{\cal{F}}}^{(d)}-{\dot{\bar{\cal{F}}}}^{(d)}{\cal{F}}^{(d)}\Big)\bigg|_{\bar{\eta}=0}\nonumber\\
&=&\frac{\partial}{\partial\bar{\eta}}\Big({\dot{{\cal{F}}}}^{(d)}{\cal{F}}^{(d)}\Big)\bigg|_{\eta = 0} \equiv \int d\bar{\eta}\;\Big({\dot{{\cal{F}}}}^{(d)}{\cal{F}}^{(d)}\Big)\bigg|_{\eta =0}\nonumber\\
    &=&\frac{\partial}{\partial\bar{\eta}}\frac{\partial}{\partial\eta}\Big({\cal{Z}}^{(d)}{\cal{F}}^{(d)}\Big) \equiv \int d\bar{\eta}\int d\eta\,\Big({\cal{Z}}^{(d)}{\cal{F}}^{(d)}\Big)\nonumber\\
    &=&-\frac{1}{2}\frac{\partial}{\partial\bar{\eta}}\frac{\partial}{\partial\eta}\Big({\Xi}^{(d)}\,{\dot{{\cal{F}}}}^{(d)}-{\dot{\Xi}}^{(d)}\,{\cal{F}}^{(d)}\Big) \equiv -\frac{1}{2}\int d\bar{\eta}\int d\eta\, \Big({\Xi}^{(d)}\,{\dot{{\cal{F}}}}^{(d)}-{\dot{\Xi}}^{(d)}\,{\cal{F}}^{(d)}\Big). \qquad
\end{eqnarray} 
It is straightforward to illustrate the nilpotency and absolute anti-commutativity of (anti-)co-BRST charges within the framework of the supervariable approach as 
\begin{eqnarray}
&&\frac{\partial}{\partial\bar{\eta}}Q_{d}= s_d Q_d = 0\;\Longleftrightarrow\;Q_{d}^{2}=0,\nonumber\\
&& \frac{\partial}{\partial{\eta}}Q_{ad} = s_{ad} Q_{ad} = 0\;\Longleftrightarrow\;Q_{ad}^{2}=0,\nonumber\\
&&\frac{\partial}{\partial\bar{\eta}}Q_{ad} = s_d Q_{ad} \;\Longleftrightarrow\;Q_{d}\,Q_{ad}+Q_{ad}\,Q_{d}=0,\nonumber\\
&& \frac{\partial}{\partial\eta}Q_{d}= s_{ad} Q_d = 0 \; \Longleftrightarrow\;Q_{d}\,Q_{ad}+Q_{ad},Q_{d}=0,
\end{eqnarray}
where we have used the nilpotency $(\partial_{\eta}^{2}=\partial_{\bar{\eta}}^{2}=0)$ and anti-commutativity $(\partial_{\eta}\partial_{\bar{\eta}}+\partial_{\bar{\eta}}\partial_{\eta}=0)$ of the Grassmannian derivatives $\partial_{\eta}$ and $\partial_{\bar{\eta}}$.

\subsection{Invariance of the  Lagrangian}
In this subsection, we consider the invariance of the Lagrangian in the language of Grassmannian translational generators $(\partial_{\eta},\partial_{\bar{\eta}})$. We begin by generalizing the first-order gauge invariant Lagrangian $L_{f}$ on to $(1,2)$-dimensional superspace as
\begin{eqnarray}
 L_{f}\longrightarrow{\cal{L}}_{f}
 &=&P_{r}\,\dot{r}+P_{\theta}\,{\dot{\Theta}}^{(h)}+P_{z}{\dot{\cal{Z}}}^{(h)}-\frac{1}{2}P_{r}^{2}-\frac{1}{2r^{2}}P_{\theta}^{2}-\frac{1}{2}P_{z}^{2}\nonumber\\
 &-&\Xi^{(h)}\big(g\,P_\theta+P_{z} \big) - V(r) = L_f.
\end{eqnarray}
This Lagrangian is independent of the Grassmannian variables, which leads to the (anti-)BRST invariance of the Lagrangian in supersace. More precisely
\begin{eqnarray}
    &&\frac{\partial}{\partial\bar{\eta}}{\cal{L}}_{f}\bigg|_{\eta = 0}=0\;\Longleftrightarrow\;s_{b}L_{f}=0,\nonumber\\
    &&\frac{\partial}{\partial{\eta}}{\cal{L}}_{f}\bigg|_{\bar \eta = 0}=0\;\Longleftrightarrow\;s_{ab}L_{f}=0.
\end{eqnarray}
Similarly, the (anti-)BRST invariant Lagrangian in $(1,2)$-dimensional superspace can be written as
\begin{equation}
    {\cal{L}}_{b}={\cal{L}}_{f}+b\,\big({\dot{\Xi}}^{(h)}-{\cal{Z}}^{(h)}\big)-\frac{1}{2}b^{2}+{\dot{\bar{\cal{F}}}}^{(h)}{\dot{\cal{F}}}^{(h)}+\bar{\cal{F}}^{(h)}{\cal{F}}^{(h)}.
\end{equation}
The quasi-invariance of the Lagrangian (${\cal{L}}_{b}$) under (anti-)BRST symmetry transformations can be captured as follows: 
\begin{eqnarray}
&&\frac{\partial}{\partial\bar{\eta}}{\cal{L}}_{b}\bigg|_{\eta=0} = \frac{d}{dt}\big(b\,{\dot{\cal{C}}}\big)\;\Longleftrightarrow\;s_{b}L_{b}=\frac{d}{dt}\big(b\,{\dot{\cal{C}}}\big),\nonumber\\
&&\frac{\partial}{\partial{\eta}}{\cal{L}}_{b}\bigg |_{\bar{\eta}=0} = \frac{d}{dt} \big(b\,{\dot{\bar{\cal{C}}}} \big)\;\Longleftrightarrow\;s_{ab}L_{b}=\frac{d}{dt}\big(b\,{\dot{\bar{\cal{C}}}} \big).
\end{eqnarray}
We represent the (anti-)BRST invariant super-Lagrangian in following different ways
\begin{eqnarray}
    {\cal{L}}_{b}&=&{\cal{L}}_{f}+\frac{\partial}{\partial\bar{\eta}}\Big[\bar{\cal{F}}^{(h)}\Big({\dot{\Xi}}^{(h)}-{\cal{Z}}^{(h)}-\frac{1}{2}\,b \Big)\Big],\nonumber\\
    &=&{\cal{L}}_{f}-\frac{\partial}{\partial{\eta}}\Big[{\cal{F}}^{(h)}\Big({\dot{\Xi}}^{(h)}-{\cal{Z}}^{(h)}-\frac{1}{2}\,b\Big)\Big].
\end{eqnarray}
This can be encoded in terms of the Grassmannian translational generators as
\begin{eqnarray}
    &&\frac{\partial}{\partial\bar{\eta}}{\cal{L}}_{b}\bigg|_{\eta=0}=0\;\Longleftrightarrow\;s_{b}L_{b}=0,\nonumber\\
    &&\frac{\partial}{\partial{\eta}}{\cal{L}}_{b}\bigg|_{\bar{\eta}=0}=0\;\Longleftrightarrow\;s_{ab}L_{b}=0,
\end{eqnarray}
where we have used the nilpotency property $(\partial_{\eta}^{2}=0,\partial_{\bar{\eta}}^{2}=0)$ of the Grassmannian derivatives.

We now demonstrate the (anti-)co-BRST invariance of the Lagrangian in the same manner. The Lagrangian given in Eq.~\eqref{ab} is generalized  to the superspace by using the supervariables given in Eq.~\eqref{sv3} as
\begin{eqnarray}
{\cal{L}}_{b}&=&P_{r}\dot{r}+P_{\theta}{\dot{\Theta}}^{(d)}+P_{z}{\dot{\cal{Z}}}^{(d)}-\frac{1}{2}P_{r}^{2}-\frac{1}{2r^{2}}P_{\theta}^{2}-\frac{1}{2}P_{z}^{2}-\Xi^{(d)}(gP_\theta+P_{z})-V(r)\nonumber\\
&+& b\big({\dot{\Xi}}^{(d)}-{\cal{Z}}^{(d)}\big)-\frac{1}{2}\,b^{2}+{\dot{\bar{\cal{F}}}}^{(d)}{\dot{\cal{F}}}^{(d)}+\bar{\cal{F}}^{(d)}{\cal{F}}^{(d)},
\end{eqnarray}
which is unaffected by the presence of Grassmannian variables in superspace. Consequently,
\begin{eqnarray}
    &&\frac{\partial}{\partial\bar{\eta}}{\cal{L}}_{b}\bigg|_{\eta=0}=\frac{d}{dt}\Big( \big(P_{z}+gP_{\theta} \big)\,{\dot{\bar{\cal{C}}}}\Big)\;\Longleftrightarrow\;s_{d}L_{b}=\frac{d}{dt}\Big(\big(P_{z}+gP_{\theta}\big)\,{\dot{\bar{\cal{C}}}}\Big),\nonumber\\
    &&\frac{\partial}{\partial{\eta}}{\cal{L}}_{b}\bigg|_{\bar{\eta}=0}=\frac{d}{dt}\Big(\big(P_{z}+gP_{\theta} \big)\,{\dot{\cal{C}}}\Big)\;\Longleftrightarrow\;s_{ad}L_{b}=\frac{d}{dt}\Big(\big(P_{z}+gP_{\theta} \big)\,{\dot{\cal{C}}}\Big).
\end{eqnarray}
which are compatible with the expressions given in \eqref{z}.

\section{Bosonic Symmetries}\label{S6}
In this section, in addition to  the fermionic symmetries, we explore the possibility of bosonic symmetries. To this end, we observed that the anti-commutator between BRST and co-BRST as well as anti-BRST 
and anti-co-BRST lead to unique novel symmetries of the system - known as bosonic symmetries. Specifically, we interpret  bosonic symmetry in terms of fermionic symmetry, as 
\begin{equation}
    \{s_{b},s_{d}\}=s_{w},\qquad \{s_{ab},s_{ad}\}=s_{\bar{w}},
\end{equation}
where $s_{w},s_{\bar{w}}$ denote the bosonic symmetries. Moreover, the following anti-commutators are identically zero 
\begin{eqnarray}
    \{s_{b},s_{ab}\}=0,\qquad \{s_{d},s_{ad}\}=0, \qquad 
     \{s_{b},s_{ad}\}=0,\qquad \{s_{d},s_{ab}\}=0.
\end{eqnarray}
Thus, the bosonic symmetry transformations, for the variables of the system, can be explicitly given as below: 
\begin{eqnarray}\label{bo}
   &&s_{w}r=0,\qquad s_{w} P_{r}=0,\qquad
    s_{w} \theta=g{\dot{b}}-g(P_{z}+gP_{\theta}),\qquad s_{w} P_{\theta}=0,\nonumber\\&& s_{w} z=\dot{b}-(P_{z}+gP_{\theta}),\qquad
     s_{w} P_{z}=0, \qquad s_{w}\zeta=b-(\dot{P}_{z}+g\dot{P_{\theta}}),\nonumber\\&&
   s_{w}{\bar{\cal{C}}}=0,\qquad s_{w}{\cal{C}}=0,\qquad s_{w} b=0, 
\end{eqnarray}
and
\begin{eqnarray}\label{abo}
   &&s_{\bar{w}}r=0,\qquad s_{\bar{w}} P_{r}=0,\qquad
    s_{\bar{w}} \theta=-g{\dot{b}}+g(P_{z}+gP_{\theta}),\qquad s_{\bar{w}} P_{\theta}=0,\nonumber\\&& s_{\bar{w}} z=-\dot{b}+(P_{z}+gP_{\theta}),\qquad
     s_{\bar{w}} P_{z}=0, \qquad s_{\bar{w}}\zeta=-b
     +(\dot{P}_{z}+g\dot{P_{\theta}}),\nonumber\\&&
   s_{\bar{w}}{\bar{\cal{C}}}=0,\qquad s_{\bar{w}}{\cal{C}}=0,\qquad s_{\bar{w}} b=0.
\end{eqnarray}
From Eqs.\eqref{bo} and \eqref{abo}, we observe that $s_{w}+s_{\bar{w}}=0$, which implies that bosonic symmetries are dependent on each other, i.e.:
\begin{equation}
    \{s_{b},s_{d}\}=s_{w}=-\{s_{ab},s_{ad}\}.
\end{equation}
The first-order Lagrangian given in Eq.\eqref{ab} remains quasi-invariant under these symmetry transformations as
\begin{eqnarray}
 s_{w}L_{b} =\frac{d}{dt}\Big(\dot{b}(P_{z}+gP_{\theta})-b(\dot{P}_{z}+g\dot{P_{\theta}})\Big) = - s_{\bar w} L_b.
\end{eqnarray}
The corresponding conserved bosonic charges, derived using Noether's theorem,
\begin{eqnarray}
Q_{w} = b^2 -\big(P_{z}+g\,P_{\theta}\big)^2 = - Q_{\bar w},
\end{eqnarray}
are generators of bosonic transformations $s_{w}$ and $s_{\bar{w}}$, respectively.

\section{Ghost Scale  and Discrete Symmetries}\label{S7}
In addition to the four fermionic and two bosonic symmetries, we observe that the Lagrangian respects another set of symmetry referred as ghost scale symmetry, as listed below
\begin{eqnarray}
    &{\cal{C}}\rightarrow e^{+1.\Lambda}{\cal{C}},\qquad \bar{{\cal{C}}}\rightarrow e^{-1.\Lambda}\bar{{\cal{C}}},\nonumber\\
    &(r,P_{r},\theta,P_{\theta},z,P_{z},\zeta,b)\rightarrow e^{0.\Lambda} (r,P_{r},\theta,P_{\theta},z,P_{z},\zeta,b),
\end{eqnarray}
where $\Lambda$ is the global independent scale parameter. The numerals in the exponential refer to the ghost number of the variables, $(\bar{\cal{C}}){\cal{C}}$ has ghost number $(-1)+1$ whereas rest of the variables have zero ghost number. Making $\Lambda$ infinitesimal, we outline the above transformations as:
\begin{eqnarray}\label{gs}
    &s_{g}{\cal{C}}={\cal{C}},\qquad s_{g}\bar{\cal{C}}=-\bar{\cal{C}}, \qquad 
    &s_{g}(r,P_{r},\theta,P_{\theta},z,P_{z},\zeta,b)=0.
\end{eqnarray}
The Lagrangian in Eq.\eqref{ab} remains unaffected by these symmetry transformations. Furthermore, we obtain the following conserved charge
\begin{equation}
Q_{g}=  \dot{\bar{\cal{C}}}\,{\cal{C}} - \bar{\cal{C}}\,\dot{\cal{C}}.
\end{equation}
It is straightforward to prove the conservation of the ghost scale charge $(Q_{g})$ using Euler-Lagrange equations of motion and the charge $Q_{g}$ is the generator of the infinitesimal continuous symmetry transformation in Eq.\eqref{gs}.

In addition, the ghost sector of the Lagrangian $L_{b}$ in Eq.\eqref{ab} also remains invariant under the following discrete symmetries
\begin{equation}
    {\cal{C}}\rightarrow \pm{\bar{\cal{C}}},\qquad {\bar{\cal{C}}}\rightarrow \mp {\cal{C}}.
\end{equation}
These discrete symmetries help us obtain the (anti-)BRST and (anti-)co-BRST transformations from BRST and co-BRST transformations respectively.

\section{Extended BRST Algebra and Cohomological Aspects}\label{S8}
So far, we have derived six continuous symmetries of the FLPR model. We now demonstrate  a connection between conserved charges corresponding to continuous symmetry transformations and de Rham cohomological operators of differential geometry \cite{9,19}. 
The conserved charges corresponding to the respective symmetries satisfy the following algebra
\begin{eqnarray}\label{coh2}
    &&Q_{(a)b}^{2}=0,\qquad Q_{(a)d}^{2}=0,\qquad \{Q_{b},Q_{ab}\}=0,\qquad\{Q_{d},Q_{ad}\}=0,\nonumber\\
    &&\{Q_{b},Q_{ad}\}=0,\qquad\{Q_{ab},Q_{d}\}=0,\qquad[Q_{g},Q_{b}]=Q_{b},\qquad[Q_{g},Q_{ab}]=-Q_{ab},\nonumber\\ 
    &&\left[Q_{g},Q_{d}\right]=-Q_{d},\qquad\left[Q_{g},Q_{ad}\right]=Q_{ad},\qquad\{Q_{b},Q_{d}\}=-\{Q_{ab},Q_{ad}\}=Q_{{w}},\nonumber\\
    &&\left[Q_{w},Q_{r}\right]=0,\qquad r=b,ab,d,ad,g,
\end{eqnarray}
here we used the canonical brackets $[r,P_{r}]=i,[\theta,P_{\theta}]=i,[z,P_{z}]=i,[\zeta,b]=i,\{{\cal{C}},{\dot{\bar{\cal{C}}}}\}=-1$ and $\{\dot{\cal{C}},\bar{\cal{C}}\}=1$.
The algebraic structures in Eq.\eqref{coh2} are redolent of the algebra of the de Rham cohomological operators of differential geometry. To be precise, these operators comprise the exterior derivative $(d)$, co-exterior derivative $(\delta)$ and Laplacian operator $(\Delta)$ and the algebra satisfied amongst them is explicitly given as \cite{20}
\begin{equation}\label{deRh}
    d^{2}=0,\qquad \delta^{2}=0,\qquad \{d,\delta\}=\Delta,\qquad \left[\Delta,d\right]=0,\qquad\left[\Delta,\delta\right]=0.
\end{equation}
It is noteworthy that the algebra given in Eq.\eqref{coh2} and \eqref{deRh} entails an exact mapping between cohomological operators and conserved charges (and their corresponding symmetry transformations). More precisely, from the nilpotency of (anti-)BRST and (anti-)co-BRST charges (their corresponding symmetry transformations), we can map them to exterior derivative or co-exterior derivative. Moreover, the anti-commutator $\{Q_{b},Q_{d}\}=Q_{w}$ is correlated to anti-commutator of the cohomological operators $\{d,\delta\}=\Delta$, which evokes the mapping of conserved charges $Q_{b},Q_{d},Q_{w}$ to the exterior derivative, co-exterior derivative and Laplacian operator, respectively. Therefore, we have the following two-to-one mapping between the conserved charges and cohomological operators: 
\begin{eqnarray}
    (Q_{b}, Q_{ad})\longrightarrow d,\qquad(Q_{ab}, Q_{d})\longrightarrow \delta,\qquad (Q_{w}, -Q_w)\longrightarrow\Delta.
\end{eqnarray}
As a result, the continuous symmetry transformations and their corresponding conserved charges furnish the physical realization of  abstract mathematical quantities  such as de Rham cohomological operators. Thus, the theory of the FLPR model brings forth a toy model for the Hodge theory.

\section{Conclusions}\label{S9}
In the present study, we have derived the off-shell nilpotent and absolutely anti-commuting (anti-)BRST and (anti-)co-BRST symmetry transformations within the framework of the supervariable approach. We generalized the one dimensional ordinary space onto $(1,2)$-dimensional superspace and exploited the horizontality condition along with the gauge invariant restrictions to derive the (anti-)BRST symmetry transformations, whereas dual horizontality condition together with the (anti-)co-BRST invariant restrictions have been exploited  to obtain the (anti-)co-BRST symmetry transformations. We have established a relation to procure the (anti-)BRST and (anti-)co-BRST symmetries of any generic variable from its corresponding supervariable in superspace. Moreover, we provided a geometrical interpretation of the key features of these symmetries such as nilpotency and anti-commutativity in terms of the translational generators $\partial_{\eta}$ and $\partial_{\bar{\eta}}$. The invariance of Lagrangian is also captured with the aid of supervariables in the superspace.

\par Furthermore, we have demonstrated that the (anti-)BRST invariant Lagrangian respects another novel symmetry known as bosonic symmetry, which is derived from the anti-commutator of (anti-)BRST and (anti-)co-BRST symmetry transformations. We also procured the ghost scale and discrete symmetry transformations and corresponding charges. Subsequently, we established that all these symmetry transformations and their generators adhere to a mathematically abstract algebra which is analogous to the algebra obeyed by the de Rham cohomological operators of differential geometry. Thus, we presented the FLPR model as a prototype for the Hodge theory.

\end{document}